# Affirmative Algorithms:
# The Legal Grounds for Fairness as Awareness[1]

Daniel E. Ho[2]    Alice Xiang[3]

## Abstract

While there has been a flurry of research in algorithmic fairness, what is less recognized is that modern antidiscrimination law may prohibit the adoption of such techniques. We make three contributions. First, we discuss how such approaches will likely be deemed "algorithmic affirmative action," posing serious legal risks of violating equal protection, particularly under the higher education jurisprudence. Such cases have increasingly turned toward anticlassification, demanding "individualized consideration" and barring formal, quantitative weights for race regardless of purpose. This case law is hence fundamentally incompatible with fairness in machine learning. Second, we argue that the government-contracting cases offer an alternative grounding for algorithmic fairness, as these cases permit explicit and quantitative race-based remedies based on historical discrimination by the actor. Third, while limited, this doctrinal approach also guides the future of algorithmic fairness, mandating that adjustments be calibrated to the entity's responsibility for historical discrimination causing present-day disparities. The contractor cases provide a legally viable path for algorithmic fairness under current constitutional doctrine but call for more research at the intersection of algorithmic fairness and causal inference to ensure that bias mitigation is tailored to specific causes and mechanisms of bias.

[1] We thank Alexandra Chouldechova, Jacob Goldin, Peter Henderson, Jessica Hwang, Mark Krass, Patrick Leahy, Laura Trice, and Chris Wan for helpful comments. Authors are listed alphabetically and have equally contributed to this work.
[2] William Benjamin Scott and Luna M. Scott Professor of Law; Professor of Political Science; Senior Fellow, Stanford Institute for Economic Policy Research; Associate Director of Stanford Institute for Human-Centered Artificial Intelligence; Director, Regulation, Evaluation, and Governance Lab, Stanford University.
[3] Head of Fairness, Transparency, and Accountability Research at the Partnership on AI.

**Introduction**

A central concern with the rise of artificial intelligence (AI) systems is bias. Whether in the form of criminal "risk assessment" tools used by judges, facial recognition technology deployed by border patrol agents, or algorithmic decision tools in benefits adjudication by welfare officials, algorithms can encode historical bias and wreak serious harm on racial, gender, and other minority groups. A substantial literature has emerged within machine learning around remedies for algorithmic bias, with the consensus position being that machine bias is best addressed through awareness of such "protected groups."[4] Responding to these concerns, the White House called for the "construction of algorithms that incorporate fairness properties into their design and execution" in 2016. The recent White House draft guidance on AI regulation similarly calls for "bias mitigation."

Yet equal-protection doctrine, particularly the affirmative-action cases, poses serious obstacles to algorithmic fairness. Equal-protection doctrine applies directly to government entities, preventing them from treating similarly situated individuals differently. The leading cases single out decision systems that formally classify and quantify by race as particularly noxious. In *Gratz v. Bollinger*, the Supreme Court found that the University of Michigan violated equal protection in awarding underrepresented minorities 20 points in a 150-point undergraduate admissions system. Michigan's practice was not narrowly tailored to achieve the compelling state interest in the educational benefits from a diverse student body. In contrast, the Supreme Court found in *Grutter v. Bollinger* that Michigan Law School's use of a "holistic" system without any formalization of points for minorities was constitutional. Narrow tailoring means "individualized consideration," which in turn "demands that race be used in a flexible, nonmechanical way."

What is less recognized is that these cases may neuter fairness in machine learning. This is because the leading approaches to remedy algorithmic bias rely on protected attributes to promote "fair" (aka unbiased) outcomes. Although many of these approaches might not ordinarily be thought of as affirmative action, just as in *Gratz*, they boil down to an adjustment that uses a protected attribute, such as race, in a kind of point system. One leading approach uses different thresholds for loan approvals for White and Black applicants to, for instance, equalize repayment rates. Another normalizes features across protected groups, such as by transforming test scores so that gender differences disappear. Yet another purposely selects features that predict the outcome, but are invariant to counterfactual changes in the protected group status. Such approaches have been packaged and billed as bias audit and mitigation measures. But anticlassification—the notion that the law should not classify individuals based on protected attributes—may render such forms of algorithmic fairness unlawful.

---

[4] We use "protected groups" to refer to groups conventionally protected under various antidiscrimination provisions, such as the Constitution's Equal Protection Clause, Title VII of the Civil Rights Act, the Fair Housing Act, and the like. An important debate concerns how such conventional classifications fail to grapple with distinct harms at the intersection of these groups. For the leading work on this, see Kimberlé Crenshaw, *Demarginalizing the Intersection of Race and Sex: A Black Feminist Critique of Antidiscrimination Doctrine, Feminist Theory and Antiracist Politics*, 1989 U. CHI. LEGAL FORUM 139.

This state of affairs is pernicious, particularly in light of two additional trends. First, machine learning has seen staggering advances over the past ten years. Use of machine learning within government has the promise to dramatically transform outdated government services, from protection of the environment to administering benefits programs to government service delivery. Second, Black Lives Matter has rightly triggered a social, intellectual, and political reckoning to grapple with longstanding forms of institutional racism.[5] Anticlassification can impede serious examination of the sources and remedies for algorithmic bias, precisely when such examination is sorely needed. As governments explore the use of AI tools, the concerns about legal risk can lead to the worst approach of all: turning a blind eye.

We consider the most viable legal justification for algorithmic fairness. In Part I, we sketch out what is known about the connection between affirmative action and algorithmic decision-making. We consider the claim that some fairness approaches do not violate anticlassification because either (a) the protected attribute is used in a holistic setting with many other contextual features, or (b) the protected attribute is used only in training, but not at deployment. While recent trends in antidiscrimination law help ascertain the legal risk of specific engineering approaches, the leading approaches are likely forms of algorithmic affirmative action, subject to strict scrutiny.

---

[5] The scholarly literature here is voluminous, but for some leading accounts on the role of government policies, see Michelle Alexander, The New Jim Crow: Mass Incarceration in the Age of Colorblindness (2010); Ira Katznelson, When Affirmative Action Was White: An Untold History of Racial Inequality in Twentieth-Century America (2006); Richard Rothstein, The Color of Law: A Forgotten History of How Our Government Segregated America (2018).

In Part II, we argue that even if algorithmic fairness is a form of affirmative action, the appropriate jurisprudence is that of government contracting, not higher education. In contrast to the fixation in the higher education cases on diversity, the contractor cases hinge on inquiries that incentivize assessment, rather than willful blindness, of bias. These inquiries turn on (a) how much discrimination is attributable to the actor, and (b) whether the design and magnitude of voluntary affirmative action to remedy such disparities can be justified through a "strong basis in evidence." This approach has key virtues: it provides the incentives to understand bias and enables the adoption of algorithmic decision tools to remedy historical bias. In Part III, we illustrate how the government-contractor cases would enable the examination of sources and remedies for algorithmic bias. In Part IV, we discuss the significant limitations of this approach, but we conclude in Part V that it remains the most promising legal justification for algorithmic fairness given the current legal landscape. Our focus in this Essay is on equal-protection doctrine, so our analysis applies primarily to government entities and may differ for private-sector entities, although much of the jurisprudence under other civil-rights provisions (e.g., Title VII of the Civil Rights Act, Equal Credit Opportunity Act) mirrors equal-protection doctrine.

## I. What Is Algorithmic Affirmative Action?

Is algorithmic fairness affirmative action? We first describe why modern antidiscrimination law is likely to consider most algorithmic fairness methods to be forms of affirmative action and also the serious legal doubt about government classification based on protected attributes. We then consider the most prominent defenses of approaches to algorithmic fairness, but conclude that there is a significant likelihood that those approaches would violate anticlassification.[6]

A. The Equal Protection Challenge

The common "group fairness" approaches take a metric of interest, such as outcomes or false-positive rates, and seek to equalize these metrics across groups. For example, one approach would vary the cutoff for a loan application to ensure that majority and minority groups have equal opportunities to secure a loan. Figure 1 illustrates this approach. The *x*-axis plots the credit score and the *y*-axis plots the number of individuals with a given credit score, with darker dots indicating defaults. The two histograms represent two demographic subgroups, which have an underlying difference in credit scores. Applying the same cutoff (dashed vertical line) would lead to different repayment rates across the subgroups. To promote equal opportunity, one algorithmic-fairness solution would be to vary the cutoff for loans by subgroup (solid vertical lines). While this has the appealing result that the repayment rates are (roughly) equalized across groups, the use of different credit score thresholds is the kind of race-based point adjustment *Gratz* deemed problematic. Such adjustments would also violate the [Equal Credit Opportunity Act](#)'s bar on race-based decisions and, in the employment context, "race-norming" of aptitude tests outlawed by the [Civil Rights Act of 1991](#). Indeed, if one converted the *x*-axis to indicate an aptitude test or a 150-point score admissions score, as in *Gratz*, it is functionally indistinguishable from affirmative action.

---

[6] For an analysis primarily under Title VII, which concludes that the most plausible rationale for algorithmic fairness is as a voluntary employer affirmative-action plan, see Jason R. Bent, *Is Algorithmic Affirmative Action Legal?*, 108 GEO. L.J. 803 (2020). That piece focuses less on the constitutional jurisprudence and we address the claim that the use of race in algorithmic-fairness models is "holistic" and therefore constitutional under *Grutter* below.

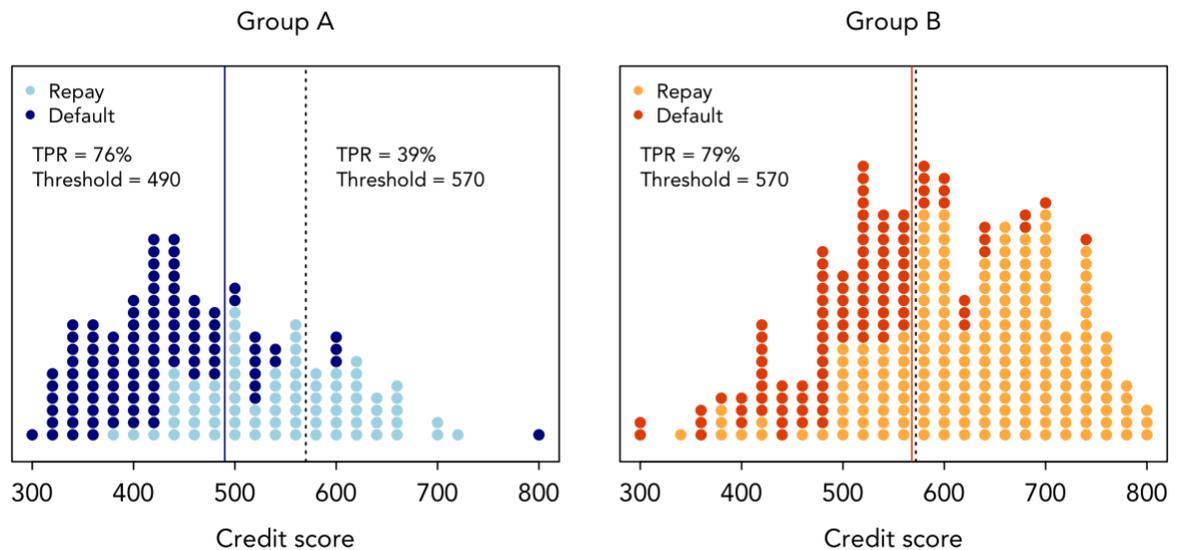

Figure 1: "Equal Opportunity" Thresholds. This figure illustrates how the "equal opportunity" bias mitigation approach,[7] which equalizes true positive rates, would lead to different thresholds between the two groups if they have different underlying distributions of credit scores. In this context, the true positive rate equals the number of individuals who would repay the loan, whose credit scores were above the threshold for approval—i.e., people who got a loan and repaid it—out of the total number of individuals who would have repaid the loan. The motivation for this approach is the idea that among individuals who would repay the loan, they should have equal opportunity to get a loan, regardless of whether they are in the orange or blue group.

But should it not be different when the protected attribute helps to predict an outcome of interest? Here, the case law offers little help. Equalization based on actuarial outcomes remains suspect. In *Craig v. Boren*, the Supreme Court, establishing intermediate scrutiny for classifications based on sex, found that a minimum drinking age that was higher for men than women violated the Equal Protection Clause. The gender difference was not substantially related to the state interest to protect public health and safety. The fact that 2 percent of men in the age bracket were arrested for driving under the influence, compared to only 0.18 percent of women, was insufficient.[8] "[P]roving broad sociological propositions by statistics is a dubious business, and one that inevitably is in tension with the normative philosophy that underlies the Equal Protection Clause." Similarly, the Supreme Court found that a pension plan cannot constitutionally charge women more, even though actuarially, women lived five years longer than men in the relevant population. "Even a true generalization about the class is an insufficient reason." In a later case, the Supreme Court also found that a pension plan may not pay women less based on differences in predicted longevity. As Justice Thurgood Marshall wrote in concurrence, "[t]he use of sex-segregated actuarial tables to calculate retirement benefits violates Title VII whether or not the tables reflect an accurate prediction of the longevity of women as a class."[9]

---

[7] Figure 1 is inspired by this visualization from Google Research.
[8] The court fumbled in its interpretation of these statistics, concluding that "a correlation of 2% must be considered an unduly tenuous 'fit.'"

Nor does it matter, under anticlassification principles, whether the goal is positive versus negative discrimination. In *Parents Involved in Community Schools v. Seattle School District No. 1*, the Supreme Court invalidated Seattle's school-assignment algorithm that used race as a tiebreaker to determine who may be admitted to an oversubscribed school. When a school was not within 10 percent of the school district's White/nonwhite racial balance, students were selected to bring the school into greater balance regardless of the direction of imbalance. Despite the seemingly benign goal of fostering integration, the Court rejected distinctions between types of classifications. An "emphasis on 'benign racial classifications' suggests . . . [the] ability to distinguish good from harmful governmental uses of racial criteria. History should teach greater humility." Famously, Chief Justice John Roberts concluded, "[t]he way to stop discrimination on the basis of race is to stop discriminating on the basis of race."

Lastly, even the possibility of creating disparate impact may be insufficient for revising a promotion mechanism. In *Ricci v. DeStefano*, the Supreme Court found that New Haven had improperly discriminated (disparate treatment) by tossing the test results of a firefighting promotion test, even though using the test results would have resulted in no Black firefighters being promoted. New Haven had made an employment decision "because of race," without a strong basis in evidence of disparate-impact liability, to which we return below.

These cases illustrate antidiscrimination law's sharp turn from antisubordination—the notion that the law should protect the

---

[9] So-called "statistical discrimination," using protected attributes to draw inferences about unobserved attributes (e.g., productivity), similarly violates antidiscrimination law's ban on making decisions because of a protected attribute. *See* David A. Strauss, *The Law and Economics of Racial Discrimination in Employment: The Case for Numerical Standards*, 79 Geo. L.J. 1619, 1623 (1991). Refusing to hire female violinists into an orchestra because ticket prices may increase due to discriminatory preferences by the audience is no defense to Title VII.

disadvantaged—toward anticlassification. Under anticlassification, the government should stop classifying based on race or gender. Period. But these developments in equal-protection jurisprudence pose a basic challenge to algorithmic fairness, which overwhelmingly relies on formalizing, classifying, and weighing predictions based on protected attributes.

B. Potential Defenses

We consider three potential defenses of algorithmic fairness approaches. First, there is the argument that such approaches use race as "just one factor among all the observed variables . . . more closely resembl[ing] the narrowly-tailored law school admissions practices upheld in *Grutter*." Under this theory, there has been no classification when the protected attribute is one of many features. The pension plans, for instance, may have been problematic because gender was the sole factor determining contribution rates. Yet under this interpretation, virtually any use of protected class attributes in the big data context, where protected attributes are inherently one of many features, might be permissible. It is highly unlikely that a court would accept this interpretation. In *Gratz*, the college also used race as only one factor in addition to high school GPA, standardized test scores, personal statement, high school quality and curriculum, residency status, alumni relationship, leadership activities, and athletics. The fact that race is one of many *potential inputs* provides no sense of its actual impact in machine learning. And indeed, bias mitigation often involves using the protected class variable not simply as a factor in the model, but in more specialized ways (e.g., as part of a fairness constraint), making the one-factor-of-many defense hard to credit.

Second, some scholars have proposed methods to "zero out" the effect of race, without using race per se in the prediction step.[10] These approaches, known as "disparate learning processes" are more appealing from a legal perspective since they make the algorithm appear more race-neutral. Professor Cynthia Dwork and colleagues, however, powerfully argue that protected attributes should not be zeroed out because they may provide important contextual value. Consider a hypothetical technology company's hiring algorithm, which predicts job performance based on candidates' resumes and transcripts, including the number of elective computer science (CS) courses they have taken. Assume that minorities have taken far fewer elective CS courses in part because they have felt alienated in the curriculum, and as a result, the number of elective CS courses is also much less predictive for minorities than for non-minorities. Training a model that interacted minority status with number of CS courses taken would lead to outcomes that are deemed "fairer" under many conceptions of fairness, while also yielding more accurate predictions. Zeroing out the effect of race would miss this important dynamic, as illustrated in Figure 2. More generally, scholars have shown that disparate learning processes can functionally result in disparate treatment and lead to suboptimal tradeoffs between fairness and accuracy. These results generally call into question whether such approaches could meet narrow tailoring.

---

[10] These approaches are sometimes referred to as "disparate learning processes."

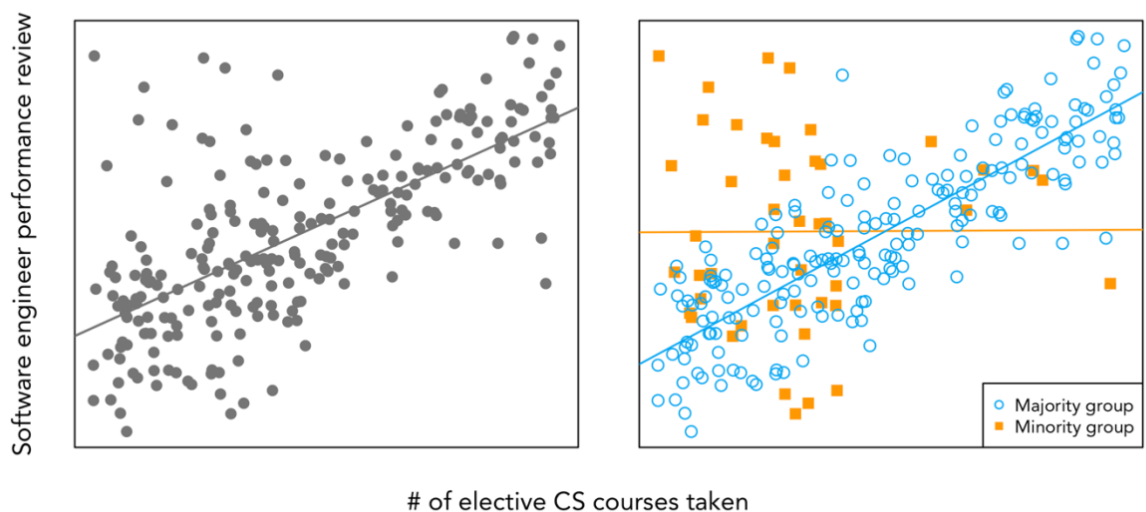

Figure 2: The Contextual Value of Protected Class Attributes. The left panel illustrates a model fitted to synthetic data when there is blindness to the protected class attribute, demonstrating a strong positive correlation between number of elective CS courses taken and performance review as a software engineer. The right panel, however, illustrates separate models: one for individuals in the majority group (blue circles) and another for the minority group (orange squares). In the majority group, there is a strong positive correlation, but in the minority group, there is virtually no correlation. The magnitude of correlation differs significantly between majority and minority groups. High-performing individuals in the minority group take fewer elective CS courses than those in the majority group, so putting significant weight on "number of elective CS courses" would result in many high performing minority candidates not being hired.

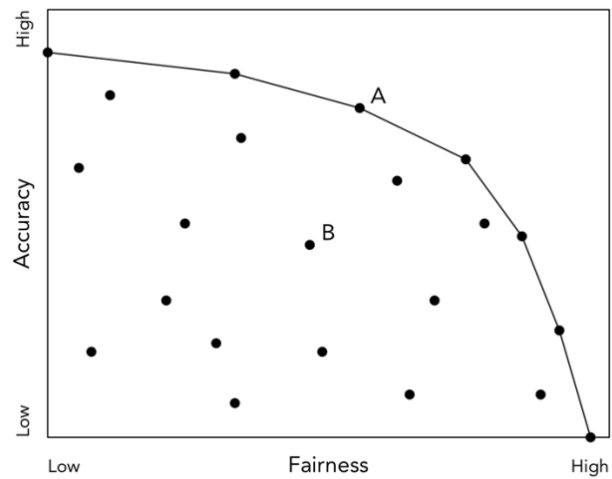

Figure 3: Pareto Frontier of the Accuracy-Fairness Trade-off. The *x*-axis presents fairness (by whatever measure) from low (left) to high (right) and the *y*-axis represents accuracy (by whatever measure) from low (bottom) to high (top). Each point indicates one potential machine learning model. The lines connect the "Pareto frontier," where there is no way to improve accuracy without degrading fairness and vice versa. Any model that is not on the Pareto frontier could improve fairness with no loss in accuracy and is hence, arguably, not the least restrictive means. In the example above about CS courses, for some accuracy and fairness metrics, the model that takes into account whether the individual is in the majority or minority group would be closer to the Pareto frontier (e.g., at point A), whereas a blind model would be in the interior (e.g., at point B).

To illustrate this potential trade-off,[11] Figure 3 plots "fairness" (based on any one of many measures) on the *x*-axis against accuracy (which can also be measured in numerous ways) on the *y*-axis. Each point in this simplified plot illustrates how specific machine-learning models trade off fairness vs. accuracy, with the line representing the "Pareto frontier." Any model in the interior space could improve accuracy without sacrificing fairness or improve fairness without sacrificing accuracy. As illustrated by the CS courses example, blindness can leave us at an interior point like B, which is strictly dominated by a model that accounts for the protected class attribute like point A. Narrow tailoring should require that a policy navigate this Pareto

---

[11] In practice, the shape of the Pareto frontier will differ depending on the dataset and specific metrics used for fairness and accuracy. The shape of the Pareto frontier in our graph was inspired in part by Figure 2 in Alexey Romanov et al., *What's in a Name? Reducing Bias in Bios Without Access to Protected Attributes*, PROC. OF NAACL-HLT 4187 (2019).

frontier, whereby any gain in accuracy would result in a loss in fairness or vice versa.

Third, another class of approaches, responding more directly to anticlassification's demands, employs proxies of the protected class variable to generate fairer outcomes. One clever approach, for instance, equalizes features across types of names, as names may encode race, without using protected attributes. Under a formalist approach, these approaches present lower legal risk,[12] but they also invite significant gaming. Moreover, if a court sees the proxy variables as simply substituting for the protected class and the approach as race-motivated, the court may still deem the approach affirmative action.[13]

In short, while legal constraints have sparked a series of machine learning solutions, there remains considerable risk that the express utilization of protected attributes—or proxies thereof—will trigger strict scrutiny and fail anticlassification.[14] On the other hand, the machine-learning literature also teaches us that alternatives that downplay the use of protected attributes will rarely meet the Pareto frontier of the accuracy-fairness tradeoff: blindness is rarely the least restrictive means.

## II. The Proper State Interest: Discrimination or Diversity

Even if algorithmic fairness is deemed affirmative action, we argue that there is an alternative doctrinal path that would make algorithmic bias mitigation legally viable.

Affirmative action was rooted in the belief that facially neutral policies would be insufficient to address past and ongoing discrimination. In his 1965 commencement address at Howard University, President

---

[12] In *Rothe Development, Inc. v. U.S. Department of Defense*, 836 F.3d 57 (D.C. Cir. 2016), the D.C. Circuit, for instance, subjected the Small Business Administration's Section 8(a) government-contracting program for "socially disadvantaged" business only to rational basis review. Despite the fact that the implementing regulation designated five racial groups as presumptively socially disadvantaged, the court concluded there was no racial classification in the statute triggering strict scrutiny on a facial challenge.

[13] *See Fisher v. Univ. of Tex.*, 136 S. Ct. 2198, 2213 (2016) (noting that Texas's Top 10 Percent Plan, which automatically admitted students in the top 10 percent of their high school class, "cannot be understood apart from its basic purpose, which is to boost minority enrollment.").

[14] A recent proposed rule from the Department of Housing and Urban Development goes so far as to create a safe harbor for algorithms that do not make use of protected class variables.

Lyndon B. Johnson delivered what has widely been lauded as "the intellectual framework for affirmative action": "You do not take a person who, for years, has been hobbled by chains and liberate him, bring him up to the starting line of a race and then say, 'You are free to compete with all the others,' and still justly believe that you have been completely fair."

While Presidents Kennedy and Johnson mandated the first forms of affirmative action in federal contracting to ameliorate historical discrimination, higher education reached a fork in the road in *Regents of the University of California v. Bakke*. Justice Powell, announcing the judgment of the Court, rejected the University of California's remedial rationale for instituting affirmative action to ameliorate discrimination. After all, UC Davis Medical School, as a relatively young institution, could not identify specific past discrimination it had engaged in. Absent such "identified discrimination," general "societal discrimination" was "an amorphous concept of injury that may be ageless in its reach into the past." Instead, Justice Powell turned to *diversity* as the legitimate compelling state interest.

Conventionally, the diversity rationale is often seen as the Court's unique permissiveness in higher education. In K-12 education, employment, and government contracting, race-conscious decision-making must be justified as a narrowly tailored remedy for historical discrimination that is traceable to the decision-making entity. Only in the context of higher education does the Court permit an alternative compelling state interest without evidence of past discrimination.

The algorithmic context, however, flips this perception on its head: diversity, construed under *Bakke* as demanding unquantifiable, individualized consideration,[15] is, if anything, *more* restrictive than historical discrimination in the algorithmic context. First, the refusal under *Gratz* to formalize any express weight to race creates unique challenges in the algorithmic context, where any preference for minorities is inherently quantifiable. Second, if narrow tailoring to achieve diversity demands "individualized consideration," algorithmic decision-making may not be possible at all.[16] In the extreme, *Gratz* and

---

[15] *See Bakke*, 438 U.S. at 319 n.53 ("So long as the university proceeds on an individualized, case-by-case basis, there is no warrant for judicial interference in the academic process.").

[16] Simply having a human in the loop is unlikely to solve this problem. The benefit of human decision-making over machine decision-making here is that the manner in which humans take into account different factors is opaque and difficult to quantify. Having a human in the loop would thus only be

*Grutter* may imply a "right to a human decision." Following the higher education case law hence leads to fundamentally incoherent results in the algorithmic decision-making context.

In contrast, the contractor cases provide a legal test that is not only compatible with, but helps to ground, algorithmic fairness. Although there have been significant challenges to these programs, the current case law permits government classifications by race to improve the share of contracts awarded to minorities under certain conditions. The critical limitation is that government must have played a role in generating outcome disparities and the program must be narrowly tailored to benefit victims of discrimination while minimizing adverse effects on others. Similarly, voluntary affirmative-action plans by employers rely on an empirical demonstration of "manifest imbalance" in the workforce, with a narrowly tailored program that does not "unnecessarily trammel" the rights of non-minorities and is temporary in duration.

Emphasized in affirmative-action cases outside of education is the benefit of encouraging entities to adopt voluntary measures. In *Johnson*, the Court concluded that "voluntary employer action can play a crucial role in . . . eliminating the effects of discrimination in the workplace, and . . . Title VII should not be read to thwart such efforts." The Court's emphasis of "the value of voluntary efforts to further the objectives of the law" is especially pertinent in the algorithmic context, where there is significant legal uncertainty around the appropriate way to address algorithmic bias.

Although presenting sufficient evidence of responsibility for historical discrimination can be challenging, as we describe next, connecting bias mitigation to evidence of discrimination enables algorithmic fairness.

### III. The Strong Basis in Evidence for Algorithmic Fairness

A. The Evidentiary Basis for Action Based on Protected Status

Key to establishing a compelling interest in remedying historical discrimination is showing a "strong basis in evidence" that the race-conscious action (or the action based on protected status) is necessary. In the contractor setting, governments must demonstrate a gap

---

beneficial if they rebalanced the algorithmic predictions in their head on the basis of protected attributes—similar to college admissions. This might make it easier to have affirmative action informed by algorithmic predictions, but it would not enable algorithmic affirmative action, whereby there is rebalancing in the data or model itself.

between the proportion of qualified, minority-owned businesses and the proportion of government contracts given to minority-owned businesses (i.e., "disparity studies"). These studies are also often coupled with qualitative evidence suggesting intentional discrimination and/or regression analyses aimed to rule out alternative explanations. This approach encourages entities to collect data on past discrimination, to empirically understand the effects of policies, and to formalize the scope of a remedial plan, making this approach better suited for the algorithmic context.

As a result, government contracting is the rare domain where the Court has permitted different thresholds by racial group or differential treatment of minority contractors. For example, many government-contracting affirmative-action programs apply a price evaluation adjustment to bids of minority contractors, so that minority contractors can win contracts even with higher bids than non-minority contractors. Similarly, programs can create set-asides for disadvantaged contractors[17] and set quantified objectives, such as having 5 percent of its contracts go to minority-owned businesses. This stands in sharp contrast to higher education, where such objectives are seen to violate *Bakke*.

The challenge defendants face in these cases is providing sufficient empirical evidence of the relevant markets and qualified contractors. Government entities have often struggled to produce specific data on minority contractor availability in their jurisdiction and to adequately distinguish between qualified and unqualified contractors, undermining the conclusions of their disparity studies. Unlike in the higher education context, where quotas and point systems are frowned upon, government-contractor cases have focused on whether the empirical evidence is (a) sufficient to show a need for remedial action and (b) whether the magnitude or form of the program is narrowly tailored based on the evidence of discrimination (e.g., whether the right demographic groups are targeted). General trends toward anticlassification have thus manifested themselves in the contracting context as higher evidentiary standards rather than prohibitions on specific types of programs. For example, although the Federal Circuit found the Department of Defense's Section 1207 program, which featured a price evaluation adjustment, unconstitutional, it did so not on the grounds that using differing thresholds are impermissible, but on the grounds that the disparity studies were insufficiently robust.[18]

---

[17] *See, e.g.*, Small Business Act § 8(a), 15 U.S.C. § 637(a) (2018). This program was challenged in *Rothe Development, Inc.*, 836 F.3d 57, but the D.C. Circuit upheld it by applying rational-basis scrutiny instead of strict scrutiny. The court reasoned that the provisions "do not on their face classify individuals by race."

B. Application to Algorithmic Context

The government-contracting line of affirmative-action cases—spanning from contractor to employment settings—provides a more coherent path forward for efforts to mitigate algorithmic bias. It provides a doctrinal grounding for technologists to develop techniques that quantify specific forms of historical discrimination and use those estimates to calibrate the extent of bias mitigation.

Consider if this approach were adopted in *Gratz*. The critical question would not be whether admissions decisions are sufficiently "individualized," but instead whether there is a strong basis in evidence that the University of Michigan's past practices were discriminatory. The university might find, for instance, that legacy preferences (or its [segregated past](), or having had a [prominent eugenicist as university president]()) caused minority applicants to be disadvantaged by an average of 20 points. If the practice of legacy preferences comes under increasing challenge, this might provide empirical grounding for the 20-point boost for minority applicants. (Harvard University's admissions files, for instance, even suggest intentional discrimination when parental connections "signify [lineage of more than usual weight]().") This approach incentivizes universities to ask the relevant empirical questions: what has been the impact of legacy preferences, and how much should we credit the claims that alumni would cease to volunteer and contribute absent a legacy admission?[19]

This approach has several other virtues in the algorithmic context. First, enabling the articulation of clear, quantifiable objectives allows technologists to take action to address bias in their tools. In most machine learning approaches, technologists have discretion to decide *how much* of a correction is desired, but little guidance for that decision. The contractor cases teach us that the correction should be calibrated to the magnitude of historical discrimination. For example, in *State v. Loomis*, the Wisconsin Supreme Court concluded that having separate risk assessment algorithms for men and women was permissible on due process grounds.[20] The data showed a large gap between crime rates for men versus women, so using a single model would significantly overestimate the probability of recidivism for women and lower accuracy overall. Yet why does that gender disparity exist? Consider the possibility that courts have been historically biased in favor of (i.e., more lenient toward) women, resulting in fewer reconvictions given the same factual circumstances (e.g., physical altercations). If so, the gender difference in recidivism rates may reflect existing gender bias by courts and a single model might actually remedy the impact of such bias. Using a risk assessment tool that classifies on gender, in contrast, could simply encode preexisting bias favoring women. Quantifying this bias would be critical to understanding (a) whether separate models should be used and (b) how large the adjustment should be.

---

[18] In fact, the Federal Circuit specifically stated that their "holding [was] grounded in the particular items of evidence offered by DOD and relied on by the district court in this case, and should not be construed as stating blanket rules, for example about the reliability of disparity studies." *Rothe Dev. Corp.*, 545 F.3d at 1049.

[19] This was the "business necessity" credited by the U.S. Department of Education's Office for Civil Rights when it investigated practices in the late 1980s and early 1990s. *See*, *e.g.*, [*The Admissions Office Strikes Back: The Process Is Fair*](), THE HARVARD CRIMSON (Nov. 26, 1990).

Second, one of the challenges in the algorithmic-fairness literature has been the dizzying set of fairness definitions, which are sometimes mutually incompatible. Understanding the source of bias can clarify which definitions, if any, are most appropriate given the historical source of discrimination. In the CS courses example above, if we did not have any knowledge about the historical sources of discrimination, we might simply use different score thresholds for different groups as a way to address the hiring gaps. Knowing that "number of CS courses" is less predictive of success as a software engineer for certain groups because of the alienation or discrimination they experienced in those classes, however, allows for a more tailored approach of putting different weights on the "number of elective CS courses" variable for different groups.[21]

Third, identifying sources of historical bias is related to the emergence of causal-inference methods in algorithmic fairness, which seek to identify the causal effect of a protected class variable and reverse this effect. These methods have the virtue of being tied closely to the causal notion of antidiscrimination law. Although current methods do not make a distinction between the specific sources of racial disparities or the historical responsibility for those disparities, they are one step in the direction of identifying and attributing discrimination reflected in the data and designing targeted interventions.

---

[20] Note that the defendant did not bring an equal-protection claim, and the court gestured at the possibility that its decision might have differed if it evaluated the claim on equal-protection grounds.

[21] We recognize that there is still a limitation to this approach, which is that the employer may bear no responsibility for the collegiate environment. We address those challenges below.

## IV. Limitations

While there are many virtues to applying the logic of the contractor cases to the algorithmic-fairness context, we acknowledge that there are key limitations.

First, as is the case for all areas of affirmative action, the [doctrine is unstable](#) and subject to significant changes in constitutional jurisprudence. A turn toward strict anticlassification in the contracting area could render algorithmic fairness constitutionally suspect. Our basic claim is simply that, given the current constitutional doctrine, the contractor cases are much more appropriate for algorithmic affirmative action. As courts grapple with the profound questions of AI, they should look to the "strong basis in evidence" standard rather than fixating on "individualized" decision making. If anything, the algorithmic fairness literature teaches us the internal incoherence of the higher education affirmative-action jurisprudence.[22]

Second, one might argue that our approach *overcorrects* for historical bias, "[unnecessarily trammel[ing]](#) the rights of other[s]." Our proposed approach, however, likely achieves the opposite—the emphasis on correcting only for empirically identifiable historical bias that is traceable to the entity in question prevents such overstepping. While *Loomis*, for instance, permitted the use of separate algorithms by sex, the contractor approach would encourage a more nuanced consideration, whereby the algorithms should be corrected for differences in how courts treat defendants by sex. Algorithmic solutions should be institutionally tailored. Fortunately, recent work in criminal justice has attempted to [isolate discretion at each stage](#) and institution, providing an empirical basis for such institutional tailoring.

Third, one of the more important objections is that the contractor approach *underprotects* the rights of minorities. A requirement to trace disparities to the entity deploying the algorithm may not capture the vast societal, governmental, and institutional sources of inequities. This fear of scoping too far, "perhaps invalidat[ing], a whole range of tax, welfare, public service, regulatory, and licensing statutes," undergirded [*Washington v. Davis*](#)'s refusal to recognize constitutional disparate-impact claims. The contractor approach might insulate entities that reorganize in a fast-moving field from responsibilities for algorithmic fairness. This is a valid concern and suggests that the case law might need to address when algorithmic outcomes can be attributed to formally distinct institutions.[23] If an agency trains a hiring algorithm with data from a technology company, biases may stem from the company's hiring practices. The agency would risk propagating those biases, but the agency would struggle to establish direct responsibility for the biases reflected in the data.

---

[22] This corroborates the points made by Ayres & Foster, [*Don't Tell, Don't Ask*](#).

In short, many biased actions might not be remedied due to the challenges of tracing historical responsibility. This limitation stems from our current antidiscrimination law, but there are some mitigating factors. A growing corpus of scholarship has shown the significant role government policy has played in creating racial disparities. For example, scholars have shown the ways in which New Deal policies acted as a form of [White affirmative action](#) and federal housing policies in the 1940s and 1950s further [segregated Black and White Americans](#). Such a record of historical discrimination may fortify the rationale for government action. As to institutional reorganizations, labor law suggests that changes in entity structure should not absolve the new entity of past liabilities. In a merger or acquisition, for instance, the resulting entity can, in certain circumstances, face [successor liability](#) for discriminatory practices of the prior entities. Moreover, in the contractor affirmative-action context, the Court has suggested an expansive view of agency responsibility. In [*Croson*](#), the Court reasoned that since states can use their spending power to remedy private discrimination, if the agency could "identif[y] [the relevant] discrimination with the particularity required by the Fourteenth Amendment," then even if they were simply a "passive participant" in a discriminatory market, they could take action to remedy it.

Fourth, another limitation lies in whether "historical" discrimination can capture the ongoing exercise of discretion in algorithmic design.[24]

---

[23] Another example is when harm is generated by multiple institutions acting simultaneously. There, attribution may become difficult.

[24] One might also worry whether historical discrimination scopes back too far. Is there something akin to a statute of limitations for historical discrimination? The strong-basis-in-evidence standard would suggest that you can go as far back as you have sufficient empirical evidence showing the connection between historical acts and present-day disparities. One additional caveat, however, comes from [*Parents Involved*](#), in which the Court discussed the history of segregation in certain school districts, but found the connection was broken by Jefferson County reaching "unitary" status after its

If an entity cannot show evidence of historical discrimination related to its own actions, it may be barred from taking proactive action. While this is an issue in both the algorithmic and non-algorithmic contexts, it especially runs counter to good public policy in the algorithmic context, where developers can more readily measure and predict, prior to deployment, whether their algorithms will exhibit biases. The case law on proactively preventing discrimination is sparse. In *Ricci v. DeStefano*, the Court ruled that New Haven had engaged in disparate treatment when it discarded the results of a firefighter promotion test to avoid disparate-impact liability. Some interpret *Ricci* to mean that bias mitigation may not be conducted after deployment of an algorithm. But as other scholars have pointed out, the case can be distinguished due to the specific reliance interests of firefighters who had spent significant time and money preparing for the exam. These reliance interests are not necessarily relevant in the algorithmic context, and bias detection and mitigation may be much more fluid than a one-time test. In addition, the Court in *Ricci* did not find any issue with the test designer's efforts to mitigate potential biases in the design of its test, including oversampling minorities in their job analyses. Disparate treatment may hence not categorically bar refining an algorithm pre- or post-deployment to address disparate impact.

Last, another objection might be that "individualized consideration" stems not from anticlassification, but from a principle of antibalkanization, namely that the government should avoid increasing group resentment. By prohibiting transparency around affirmative action, *Gratz* and *Grutter* may have reduced racial resentment, as applicants cannot individually identify whether they were harmed by the policy. Similarly, *Ricci* may be motivated by the racial resentment of White firefighters, the "visible victims" who had spent months preparing for the exam. Antibalkanization could ultimately imply the right to a human decision when implementing affirmative action, gutting substantial potential gains—including an understanding of bias—from AI adoption. Antibalkanization, however, does not appear to explain the express use of quantitative affirmative-action targets in government contracting.

## V. Conclusion

At a time when there are growing calls for action to address algorithmic bias, machine learning and antidiscrimination law appear

---

desegregation decree was dissolved. This suggests a limitation based on the degree of empirical evidence and legal liability.

to be at an impasse. The irony of modern antidiscrimination law is that it may *prevent* the active mitigation of algorithmic bias. Current legal uncertainty has led to a pernicious state of affairs: agencies and technologists are reluctant to assess and mitigate bias in algorithmic systems. We have attempted to sketch an alternative path to break this impasse, and now conclude with three final thoughts.

First, the lesson for the machine-learning community is that the future of algorithmic fairness lies in tying bias mitigation to historical discrimination that can be empirically documented and attributable to the deploying entity. This requires a closer collaboration with social scientists to understand institutional sources of bias and with lawyers to align solutions with legal precedent. If such work is not conducted, algorithmic fairness may reach a legal dead end.

Second, algorithmic decision-making has broad lessons for the law as well. Recent results demonstrate that race- or gender-neutral approaches are less effective at promoting the state interests related to fairness and accuracy (i.e., not on the Pareto frontier).[25] And machine learning highlights the internal incoherence of narrow tailoring in higher education: how should we ensure that a race-conscious plan is the least restrictive alternative without some formalization of the weight placed on race? While the government-contracting cases also present challenges, they do not present the same fundamental incoherence. Instead, they incentivize (a) collecting empirical evidence to assess past discrimination and (b) calibrating bias mitigation to that evidence. What might be a higher evidentiary burden in the non-algorithmic context can be a lighter one in the algorithmic context.

Third, while we find the status quo of inaction due to legal uncertainty concerning, we also do not take lightly the use of protected-class variables. In medicine, for instance, there has been a long history of using race as a variable in ways that have likely amplified rather than mitigated bias, often based on egregious misconceptions about race. The idea that race has a genetic basis has made the use of race variables more palatable in clinical diagnostic tools, but this has led to misattribution of health outcome differences to genetic rather than social sources. A similar logic has carried over into the insurance context, where states do not uniformly prohibit the use of race in setting premiums, and to the calculation of economic loss in torts.[26]

---

[25] *Grutter*, 539 U.S. at 339 ("Narrow tailoring does, however, require serious, good faith consideration of workable race-neutral alternatives").
[26] *See McMillan v. City of New York*, 253 F.R.D. 247 (E.D.N.Y. 2008); Martha Chamallas, *Civil Rights in Ordinary Tort Cases: Race, Gender, and the Calculation of Economic Loss*, 38 LOY. L.A. L. REV. 1435 (2005).

The anticlassification approach posits that there is no cognizable distinction between benign and invidious government racial classifications.[27] One virtue of the contractor jurisprudence is that it utilizes race-consciousness first to assess past discrimination; with a clearer understanding of the mechanisms of discrimination, it may be easier to determine the basis, scope, and necessary tailoring for fairness as awareness.

While not without limitations given current law, redirecting attention to the government-contracting line of affirmative-action cases thus may be a viable path for algorithmic innovation and fairness.

---

[27] In *Adarand Constructors, Inc. v. Peña*, 515 U.S. 200, 225 (1995), the Court decided that all racial classifications are subject to strict scrutiny, regardless of whether they are "benign." Although subsequent courts have interpreted this as meaning that there are no "benign" uses of racial classification, the *Adarand* Court was actually motivated not by the idea that there are no benign uses, but rather, that strict scrutiny is necessary to make this distinction in practice.